\documentclass[iop]{emulateapj}

\usepackage{times,natbib,graphicx,amsmath}

\newcommand{\nustar}{\textit{NuSTAR}}
\newcommand{\nicer}{\textit{NICER}}
\newcommand{\xmm}{{\it XMM-Newton}}

\newcommand{\fluxcgs}{ergs~s$^{-1}$~cm$^{-2}$}

\newcommand{\rin}{$R_{in}$}

\shorttitle{{\it NICER} Detects Multiple Features in Ser X-1}
\shortauthors{Ludlam et al.}

\begin{document}

\title{Detection of Reflection Features in the Neutron Star Low-Mass X-ray Binary Serpens X-1 with {\it NICER}}
\author{R. M. Ludlam\altaffilmark{1},
J. M. Miller\altaffilmark{1},
Z. Arzoumanian\altaffilmark{2},
P. M. Bult\altaffilmark{2},
E. M. Cackett\altaffilmark{3},
D. Chakrabarty\altaffilmark{4},
T. Dauser\altaffilmark{5}, 
T. Enoto\altaffilmark{6}, 
A. C. Fabian\altaffilmark{7},
J. A. Garc\'{i}a\altaffilmark{8},  
K. C. Gendreau\altaffilmark{2}, 
S. Guillot\altaffilmark{9,10},
J. Homan\altaffilmark{11,12},
G. K. Jaisawal\altaffilmark{13},
L. Keek\altaffilmark{14},
B. La Marr\altaffilmark{4}, 
C. Malacaria\altaffilmark{15,16,17},
C. B. Markwardt\altaffilmark{2},  
J. F. Steiner\altaffilmark{4}, 
T. E. Strohmayer\altaffilmark{2}
}
\altaffiltext{1}{Department of Astronomy, University of Michigan, 1085 South University Ave, Ann Arbor, MI 48109-1107, USA}
\altaffiltext{2}{X-ray Astrophysics Laboratory, Astrophysics Science Division,
NASA/GSFC, Greenbelt, MD 20771, USA}
\altaffiltext{3}{Department of Physics \& Astronomy, Wayne State University, 666 W. Hancock St., Detroit, MI 48201, USA}
\altaffiltext{4}{MIT Kavli Institute for Astrophysics and Space Research, Massachusetts Institute of Technology, Cambridge, MA 02139, USA}
\altaffiltext{5}{Remeis Observatory \& ECAP, Universit\"{a}t Erlangen-N\"{u}rnberg, Sternwartstr. 7, 96049, Bamberg, Germany}
\altaffiltext{6}{Department of Astronomy, Kitashirakawa-Oiwake-cho, Sakyo-ku, Kyoto, 606-8502, Japan}
\altaffiltext{7}{Institute of Astronomy, Madingley Road, Cambridge CB3 0HA, UK}
\altaffiltext{8}{Cahill Center for Astronomy and Astrophysics, California Institute of Technology, Pasadena, CA 91125}
\altaffiltext{9}{CNRS, IRAP, 9 avenue du Colonel Roche, BP 44346, F-31028 Toulouse Cedex 4, France}
\altaffiltext{10}{Universit\'{e} de Toulouse, CNES, UPS-OMP, F-31028 Toulouse, France}
\altaffiltext{11}{Eureka Scientific, Inc., 2452 Delmer Street, Oakland, CA 94602, USA}
\altaffiltext{12}{SRON, Netherlands Institute for Space Research, Sorbonnelaan 2, 3584 CA Utrecht, The Netherlands}
\altaffiltext{13}{National Space Institute, Technical University of Denmark, Elektrovej 327-328, DK-2800 Lyngby, Denmark}
\altaffiltext{14}{Department of Astronomy, University of Maryland, College Park, MD 20742, USA}
\altaffiltext{15}{NASA Marshall Space Flight Center, NSSTC, 320 Sparkman Drive, Huntsville, AL 35805, USA}\altaffiltext{16}{NASA Postdoctoral Fellow}
\altaffiltext{17}{Universities Space Research Association, NSSTC, 320 Sparkman Drive, Huntsville, AL 35805, USA}

\begin{abstract} 
We present {\it Neutron Star Interior Composition Explorer} (\nicer) observations of the neutron star low-mass X-ray binary \mbox{Serpens~X-1} during the early mission phase in 2017. With the high spectral sensitivity and low-energy X-ray passband of {\it NICER}, we are able to detect the Fe~L line complex in addition to the signature broad, asymmetric Fe~K line. We confirm the presence of these lines by comparing the {\it NICER} data to archival observations with {\it XMM-Newton}/RGS and {\it NuSTAR}. Both features originate close to the innermost stable circular orbit (ISCO). When modeling the lines with the relativistic line model {\sc relline}, we find the Fe~L blend requires an inner disk radius of $1.4_{-0.1}^{+0.2}$ $R_{\mathrm{ISCO}}$ and Fe~K is at $1.03_{-0.03}^{+0.13}$ $R_{\mathrm{ISCO}}$ (errors quoted at 90\%). This corresponds to a position of $17.3_{-1.2}^{+2.5}$ km and $12.7_{-0.4}^{+1.6}$ km for a canonical neutron star mass ($M_{\mathrm{NS}}=1.4\ M_{\odot}$) and dimensionless spin value of $a=0$. Additionally, we employ a new version of the {\sc relxill} model tailored for neutron stars and determine that these features arise from a dense disk and supersolar Fe abundance.  
\end{abstract}


\section{Introduction}
In low-mass X-ray binary (LMXB) systems, where the companion star has mass $\leq1\ \mathrm{M}_{\odot}$,  accretion onto the compact object generally occurs through an accretion disk formed via Roche-lobe overflow. In many instances, these disks are illuminated by hard X-rays coming either from a hot electron corona  \citep{sunyaev91} or the surface of the neutron star or boundary layer (where the material from the disk reaches the neutron star; \citealt{PS01}).  
The exact location and geometry of the corona is not known, but is considered to be compact and close to the compact object (see \citealt{degenaar18} for a review and references therein). 
Regardless of the source of the hard X-rays, the disk reprocesses the illuminating photons and re-emits them in a continuum with a series of atomic features and Compton backscattering hump superimposed, known as the \lq \lq reflection" spectrum. The most prominent feature that arises as a result of reflection is the Fe~K emission line between 6.4--6.97 keV. 
The entire Fe line profile is shaped by strong Doppler and relativistic effects due to the disk's rotational velocity and proximity to the compact object \citep{Fabian89}. The extent of the red wing thereby enables important physical insights to be derived from these systems. Moreover, the blue-shifted emission of the Fe line profile provides an indication of the inclination of the disk due to Doppler effects becoming more prominent with increasing inclination \citep{dauser10}.
%The extent of the red wing is shaped by strong Doppler and relativistic effects due to the disk's proximity to the compact object \citep{Fabian89}, thereby enabling important physical insights to be derived from these systems.  
This feature has been reported in both black hole (BH: e.g., \citealt{miller02}) and neutron star (NS: e.g., \citealt{bs07}; \citealt{cackett08}) LMXBs, suggesting similar accretion geometries despite the mass difference of the compact accretor and the presence of a surface. 

An additional prominent reflection feature that can arise from the illuminated accretion disk is the lower-energy Fe~L line near $1$ keV. This feature was first reported in \citet{Fabian09} for the active galactic nucleus (AGN) 1H0707$-$495 with the same asymmetric broadening seen in Fe~K. Moreover, the ratio of Fe~K to Fe~L emission was consistent with predictions from atomic physics. 
This feature was soon discovered in other AGN, such as IRAS~13224$-$3809 (\citealt{ponti10}), cementing the importance of reflection features in these accreting systems. 

The BHs in LMXBs are scaled-down versions of the much more massive accretors in AGN \citep{miller07}. Since accretion in BH and NS LMXBs is similar, we expect to find an Fe~L feature in a NS LMXB if the conditions are right. 
There have been a number of reports of line complexes near $\sim1$ keV in NS LMXBs during persistent emission that have been attributed to the Fe~L transition, but also K-shell transitions of medium-Z elements (\citealt{vrtilek88}; \citealt{kuulkers97}; \citealt{schulz99}; \citealt{sidoli01}; \citealt{cackett10}). These lines appear to be broadened by the same mechanism as the Fe K component \citep{Ng10} and can be modeled as smeared relativistic  lines \citep{iaria09}.

Serpens~X-1 (Ser~X-1) is an \lq \lq atoll" NS LMXB located at a distance of $7.7\pm0.9$ kpc \citep{galloway08}. Optical spectroscopy and some X-ray reflection studies indicate that the system has a low binary inclination ($i\leq10^{\circ}$, \citealt{cornelisse13}; \citealt{miller13}), though higher inclinations have been reported from other X-ray reflection studies ($25^{\circ}<i<50^{\circ}$, \citealt{cackett08}, \citeyear{cackett10}; \citealt{chiang16}; \citealt{matranga17}). The low amount of absorbing material in the direction of \mbox{Ser X-1}, as demonstrated by the low neutral hydrogen column density ($\mathrm{N}_{\mathrm{H}}=4\times10^{21}$ cm$^{-2}$, \citealt{dl90}), provides an opportunity to detect multiple reflection features. 

With the recent launch of the {\it Neutron Star Interior Composition Explorer} ({\it NICER}; \citealt{gendreau}), we now have the opportunity to test reflection predictions and probe the innermost region of the accretion disk in \mbox{Ser X-1}. {\it NICER} was installed on the International Space Station in 2017 June.
The payload comprises 56 \lq \lq concentrator" optics that each focus X-rays in the 0.2--12 keV range onto a paired silicon drift detector. Prelaunch testing left 52 functioning detectors providing a total collecting area of $1900$ cm$^{2}$ at 1.5 keV with which to search for low-energy reflection features.

\setcounter{footnote}{0}

\section{Observations and Data Reduction}
The following subsections detail the reduction of \mbox{Ser X-1} observations obtained with {\it NICER}, {\it NuSTAR}, and {\it XMM-Newton}. The {\it NuSTAR} and {\it XMM-Newton} data were not acquired contemporaneously with our \nicer\ observations, but are used as a baseline for determining which features are astrophysical in the {\it NICER} data, since \mbox{Ser X-1} has remained roughly steady in its persistent emission (0.2--0.3 Crab) in the {\it Swift}/BAT and MAXI wide-field monitors for the past decade. 

\subsection{\it NICER}
{\it NICER} observed \mbox{Ser X-1} thirteen times between 2017 July and 2017 November (ObsIDs 1050320101--1050320113) for a cumulative exposure of 39.9 ks on target.  The data were reduced using {\sc NICERDAS}  version 2018-02-22\_V002d. Good time intervals (GTIs) were created using {\sc nimaketime} selecting COR\_SAX$\geq$4, to remove high particle radiation intervals associated with the Earth's auroral zones, and separating orbit day (SUNSHINE==1) from orbit night (SUNSHINE==0) in addition to the standard {\it NICER} filtering criteria. Moreover, we only selected events that occurred when the angle between the Sun and target of observation were  $\geq90^{\circ}$.
These GTIs were applied to the data via {\sc niextract-events} selecting events with PI channel between 25 and 1200 (0.25--12.0 keV) that triggered the detector readout system's slow and, optionally, fast signal chains. 
Background spectra were created from data acquired from one of seven \lq \lq blank sky" targets based on RXTE background fields \citep{jahoda06}. We reduced all observations of the background fields as described above. \mbox{Ser X-1} is much brighter (1597 counts s$^{-1}$) in comparison to the background fields ( $\sim0.7-2.3$ counts s$^{-1}$). We proceeded with using RXTE background field 5 throughout the remaining analysis since the results are not dependent upon this choice. 

The resulting event files were read into {\sc xselect} and combined to create light-curves and time-averaged spectra for orbit day and orbit night. There were no Type-I X-ray bursts present in the light-curves during the GTIs so no additional filtering was needed. The spectra suffered from instrumental residuals given the preliminary calibration at this stage. In order to mitigate these residuals we normalized the data to {\it NICER} observations of the Crab Nebula, which has a featureless absorbed power-law spectrum in the energy range of interest (see, e.g., \citealt{weisskopf10}). 

We use ObsIDs 1011010101, 1011010201, and 1013010101-1013010123 for the Crab, and the same data reduction procedure as above. The resulting exposure time for the Crab is $\sim1.1$ ks for orbit day and $\sim21.2$ ks for orbit night. The time-averaged Crab spectrum was fit with an absorbed power-law model from 0.25--10 keV to determine the absorption column along the line of sight. The absorption column was consistent with the \citet{dl90} value at $\mathrm{N}_{\mathrm{H}}\sim3.8\times10^{21}$ cm$^{-2}$. We then froze the absorption column and fit the 3--10 keV subset of the spectrum to prevent instrumental features at low energies from skewing the fit. This returned a photon index of $\Gamma\sim2.0$\footnote{We verified that the choice of background did not change the photon index by performing fits with each field. The photon index changes by no more than $\Delta\Gamma=0.005$.}. We extrapolated the fit back down to 0.25 keV and created a fake spectrum using the \lq fakeit' command within the {\sc xspec} software package \citep{arnaud96} for the same exposure time as the actual Crab data. We used the {\sc ftool} {\sc mathpha} to divide the Crab spectrum by the simulated Crab data. This yielded a spectrum with just the instrumental residuals. We then used {\sc mathpha} to divide the \mbox{Ser X-1} count rate spectrum by the instrumental residual spectrum. The normalized spectra after applying all of our filtering criteria for orbit night and day have exposures of $\sim4.5$ ks and $\sim4.7$ ks, respectively. However, since the exposure time for the Crab during orbit day was much smaller in comparison to orbit night, it introduced noise into the \mbox{Ser X-1} spectrum when normalizing. We therefore only focused on the data that was accumulated during orbit night. See Figure 1 for a comparison of the \mbox{Ser X-1} data before and after normalizing to the Crab. The source spectrum was grouped via  {\sc grppha} to have a minimum of 25 counts per bin.

\begin{figure}
\begin{center}
\includegraphics[width=8.8cm]{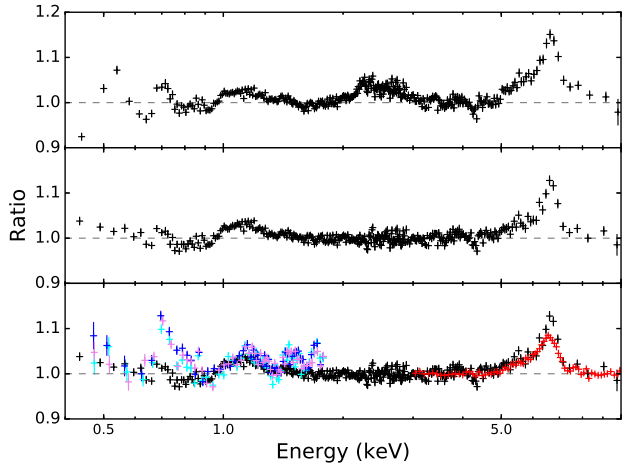}
\caption{{\it Top:} Ratio of \nicer\ data to continuum before normalizing to the Crab. Instrumental residuals can be seen at $\sim0.5$ keV, $\sim1.8$ keV, and $\sim2.2$ keV due to O,  Si, and Au, respectively. The feature at $\sim0.7$ keV has instrumental origin as well, likely due to gain offsets as the response changes rapidly in this area. {\it Middle:} Ratio of \nicer\ data to continuum after normalizing to the Crab to mitigate instrumental residuals. Emission features are still present near $1.1$ keV and $6.7$ keV. {\it Bottom:} Ratio of the data to continuum model for \nicer\ (black), \nustar\ (red), and {\it XMM-Newton}/RGS (blue, cyan, purple) observations. 
The data were fit with the combination of an absorbed single-temperature blackbody, multi-temperature blackbody, and power-law. The $0.9-1.3$ keV and $5.0-8.0$ keV energy bands were ignored while fitting to prevent the lines from skewing the continuum. Continuum parameters were tied between \nicer\ and \nustar\ while normalization components were allowed to vary. The \xmm\ data were left free with a constant allowed to float between observations. 
Data were rebinned for clarity.}
\label{fig:Felines}
\end{center}
\end{figure}

\subsection{\it NuSTAR} 
Two observations of \mbox{Ser X-1} were taken with {\it NuSTAR} on 2013 July 12 and 13 (ObsIDs 30001013002 and 30001013004) for $\sim30.5$ ks. These observations have been previously reported by \citet{miller13} and \citet{matranga17}. There were no Type-I X-ray bursts that occurred during the 2013 observations. Using the {\sc{nuproducts}} tool from {\sc nustardas} v1.8.0 with {\sc caldb} 20180126, we created light-curves and spectra for the 2013 observations with {\sc statusexpr}=\lq\lq STATUS==b0000xxx00xxxx000" to correct for high count rates. We used a circular extraction region with a radius of 100$^{\prime \prime}$ centered around the source to produce a source spectrum for both the FPMA and FPMB. We used another 100$^{\prime \prime}$ radial region away from the source for the purpose of background subtraction. Following \citet{miller13}, we respectively combined the two source spectra, background spectra, ancillary response matrices and redistribution matrix files via {\sc addascaspec} and {\sc addrmf}, weighting by exposure time. The combined spectra were grouped  to have a minimum of 100 counts per bin using {\sc grppha}. 

\begin{figure}
\begin{center}
\includegraphics[width=9.2cm]{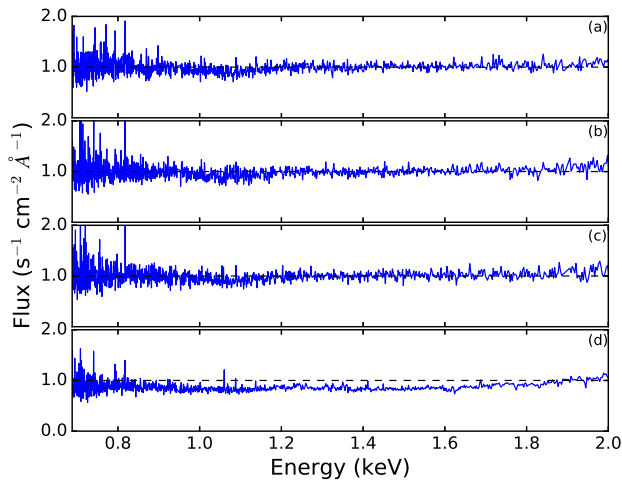}
\caption{Ratio of RGS first to second order fluxed spectra for the three {\it XMM-Newton} observations: (a) 0084020401, (b) 0084020501, and (c) 0084020601. The bottom panel denoted (d) shows an observation of GRO J1655-40 that is piled-up. The ratio is roughly consistent with unity in panels (a)-(c) indicating that pile-up in the RGS band is not an issue for Serpens X-1.}
\label{fig:pile}
\end{center}
\end{figure}

\subsection{XMM-Newton}
There are three observations of \mbox{Ser X-1} with {\it XMM-Newton} (ObsIDs 0084020401, 0084020501, and 0084020601) performed in 2004 March for a total exposure of $\sim65.7\ \mathrm{ks}$. These observations were reported in \citet{bs07}, \citet{cackett10}, and \citet{matranga17}. We focus on the Reflection Grating Spectrometer (RGS) data because we are interested in the high-resolution, low-energy spectral features. The data were reduced using the command {\sc rgsproc} in SAS v16.1. We checked that the data do not suffer from pile-up by inspecting the ratio of the first and second order fluxed spectra\footnote{XMM-Newton Users Handbook \S 3.4.4.8.3} (see Figure 2). Each ratio for \mbox{Ser X-1} is consistent with unity across nearly all of the RGS energy band indicating that pile-up is not an issue. For comparison, we plot the ratio of the fluxed spectra for the stellar mass black hole X-ray binary GRO J1655-40, which suffered from pile-up in the RGS instrument, in the bottom panel. Since the observations of \mbox{Ser X-1} were not piled-up, the first order RGS1 and RGS2 data were combined via {\sc rgscombine} for each respective observation. The resulting spectra were then grouped using {\sc grppha} to have a minimum of 25 counts per bin.  

\section{Spectral Analysis and Results}
We use {\sc xspec} version 12.9.1m and report uncertainties at the 90\% confidence level. Since the \xmm\ and \nustar\ data have been previously analyzed and published elsewhere, we choose to mainly focus on the \nicer\ results. We model the \nicer\ data in the $0.4-10.0$ keV energy band, outside of which the effective area drops sharply. We model the absorption along the line of sight using {\sc tbnew}\footnote{http://pulsar.sternwarte.uni-erlangen.de/wilms/research/tbabs/} with {\sc vern} cross sections \citep{vern96} and {\sc wilm} abundances \citep{wilms00}. We allow the neutral hydrogen absorption, as well as the oxygen and iron absorption abundances, to be free parameters to ensure that edges in the region of interest are properly modeled.  Allowing the oxygen and iron absorption to deviate from solar abundance provides a statistical improvement in the following fits at more than $9\sigma$ confidence level in each case, although these could be instrumental in origin as the O~K edge lines up with changes in the effective area of the detector and may not reflect actual ISM abundance measurements. We find a neutral absorption column of $\mathrm{N}_{\mathrm{H}}\sim7\times10^{21}$ cm$^{-2}$, which is higher than the \citet{dl90} value of $4\times10^{21}$ cm$^{-2}$, but consistent with other values reported when fitting low-energy X-ray data from {\it Chandra} and \xmm\ (\citealt{chiang16}; \citealt{matranga17}). 

We apply the double thermal-continuum model for atoll sources in the soft state to the {\it NICER} data. This model consists of a multi-temperature blackbody component ({\sc diskbb}) to model the disk emission and a single temperature blackbody component ({\sc bbody}) to model emission originating from the surface of the neutron star or boundary layer. This provides a poor fit, with  $\chi^{2}/d.o.f. = 1891.91/952$. We proceed with adding a power-law component, which is sometimes needed in the soft state \citep{lin07} and has previously been reported for \mbox{Ser X-1} (\citealt{miller13};  \citealt{chiang16}). The power-law component produced a photon index of $\Gamma=2.88\pm0.25$ and normalization of $0.69\pm0.08$ photons keV$^{-1}$ cm$^{-2}$ s$^{-1}$ at 1 keV. The overall fit improves by $10\sigma$ ($\chi^{2}/d.o.f. = 1671.33/950$) although it is still poor due to the presence of emission features. The disk component yielded a temperature of $1.07\pm0.03$ keV and normalization of $154\pm17$ km$^{2}$/ (D/10 kpc)$^{2}$ cos($i$). The single-temperature blackbody emerged with a temperature of $1.75\pm0.03$ keV and normalization $K=5.5\pm0.02\times10^{-2}$. Replacing the single temperature blackbody component for thermal Comptonization ({\sc nthcomp}) did not improve the fit ($\chi^{2}/d.o.f. = 1862.39/949$).

\begin{figure}
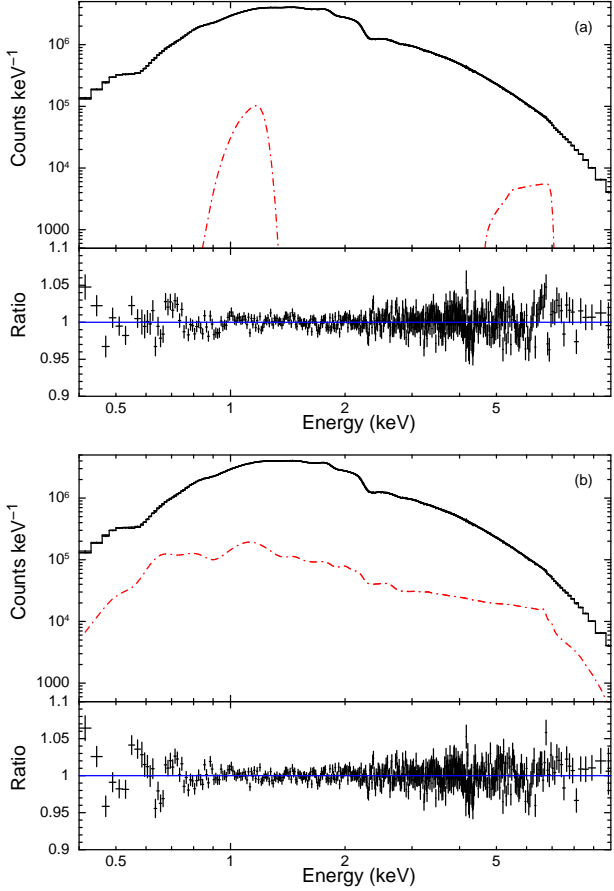

\begin{center}
\includegraphics[width=6.0cm, angle=270]{relline_pllr.eps}
\includegraphics[width=6.0cm, angle=270]{relxillns_pllr.eps}
\caption{Spectra of Serpens~X-1 with reflection modeled using {\sc relline} (a) to account for the Fe~L and K lines individually and {\sc relxillNS} (b) to account for entire reflection spectrum indicated by the red dot-dashed lines. The lower panels indicate the ratio of the \nicer\ data to overall model. The continuum is modeled with an absorbed disk blackbody, single temperature blackbody, and power-law component. For plotting purposes, the data were rebinned 
%The bottom figure (c) indicates the unfolded reflection model.
}
\label{fig:Felines}
\end{center}
\end{figure}

There are two strong emission features near $1.1$ and $6.7$ keV that can be attributed to a blend of Fe~L and Fe~K shell emission (see Figure 1). As a consistency check, we overplot the \nustar\ and \xmm\ ratio-to-continuum spectra in the lower panel of Figure 1. The presence of these features in other detectors verifies that they are not due to the \nicer\ instrumentation.  We initially apply Gaussian profiles to each feature, which  improves the fit by $\Delta\chi^{2} = 290$ for six degrees of freedom. The low-energy emission feature has a line centroid energy of $1.12\pm0.01$ keV with width $\sigma=0.10_{-0.01}^{+0.02}$ keV and normalization of $9.8\pm0.2\times10^{-3}$ photons cm$^{-2}$ s$^{-1}$, which is similar to the values reported in \citet{cackett10} for the \xmm/PN data. We again checked that the low-energy feature is not an instrumental artifact by fixing the width of the line to 0,  which is a delta function in {\sc xspec} and indicates the resolution of the detector. We find that a line width of zero is ruled out at $9\sigma$, corroborating that the line is not native to the instrumentation. The Fe~K emission feature has a line centroid energy of $6.59\pm0.04$ keV with width $\sigma=0.26_{-0.05}^{+0.06}$ keV and normalization of $3.8\pm0.1\times10^{-3}$ photons cm$^{-2}$ s$^{-1}$. The equivalent widths of the Fe~K and Fe~L lines are $0.064\pm0.002$ keV and $0.0090\pm0.0001$ keV, respectively.  The equivalent width of the Fe~K line is comparable to values reported in \citet{cackett10} for \xmm\ and {\it Suzaku} observations of other NS LMXBs, whereas the Fe L blend agrees with \citet{vrtilek88}.

\begin{table}
\caption{Reflection Modeling of Ser X-1}
\label{tab:model} %relline_nicer.xcm, relxillns_030418.xcm
\begin{center}
\begin{tabular}{lccc}
\hline
Model & Parameter & {\sc relline} & {\sc relxillNS}\\
\hline
{\sc tbnew} &$\mathrm{N}_{\mathrm{H}}$ ($10^{21}$ cm$^{-2}$)&$6.9\pm0.1$&$6.38\pm0.02$\\
&$A_{O}$& $1.10\pm0.01$&$1.25_{-0.05}^{+0.02}$\\
&$A_{Fe}$& $1.12_{-0.02}^{+0.09}$&$0.88_{-0.04}^{+0.09}$\\
{\sc diskbb}& kT (keV) & $1.13_{-0.02}^{+0.03}$&$1.15_{-0.02}^{+0.05}$\\
&norm &$130_{-20}^{+10}$&$120_{-10}^{+3}$\\
{\sc bbody} & kT (keV) &$1.80\pm0.02$&$1.85_{-0.03}^{+0.11}$\\
& K ($10^{-2}$)& $4.9\pm0.2$&$3.1_{-0.4}^{+0.6}$\\
{\sc powerlaw}& $\Gamma$& $2.5\pm0.1$&$1.8_{-0.1}^{+0.5}$\\
&norm ($10^{-1}$)& $5.0_{-0.4}^{+0.7}$&$3.50\pm0.04$\\
{\sc relline}$_{1}$&LineE (keV) &$6.96_{-0.07}^{+0.01}$& ...\\
&$q$&$3.00_{-0.16}^{+0.08}$&...\\
&$i$ ($^{\circ}$)&$10.1_{-0.2}^{+4.9}$&...\\
&$R_{in}$ ($R_{\mathrm{ISCO}}$)&$1.03_{-0.03}^{+0.13}$&...\\
&$R_{in}$ ($R_{g}$)&$6.2_{-0.2}^{+0.8}$&...\\
&$R_{in}$ (km)&$12.7_{-0.4}^{+1.6}$&...\\
&norm ($10^{-3}$)&$7.1_{-1.0}^{+0.9}$&...\\
{\sc relline}$_{2}$&LineE (keV)&$1.22_{-0.02}^{+0.01}$&...\\
&$R_{in}$ ($R_{\mathrm{ISCO}}$)&$1.4_{-0.1}^{+0.2}$&...\\
&$R_{in}$ ($R_{g}$)&$8.4_{-0.6}^{+1.2}$&...\\
&$R_{in}$ (km)&$17.3_{-1.2}^{+2.5}$&...\\
&norm ($10^{-3}$)&$6.8_{-0.7}^{+1.2}$&...\\
{\sc relxillNS}&$q$ & ...& $2.51_{-0.17}^{+0.04}$\\
&$i$ ($^{\circ}$)& ...&$4.43_{-0.30}^{+0.01}$\\
&$R_{in}$ ($R_{\mathrm{ISCO}}$)&...&$1.18_{-0.02}^{+0.10}$\\
&$R_{in}$ ($R_{g}$)&...&$7.08_{-0.12}^{+0.60}$\\
&$R_{in}$ (km)&...&$14.6_{-0.3}^{+1.2}$\\
&$\log \xi$&...&$3.2\pm0.1$\\ 
&$A_{Fe}$&...&$4.8_{-0.3}^{+0.8}$\\
&$\log N$ (cm$^{-3}$)&...&$18.83\pm0.04$\\
&$f_{\mathrm{refl}}$ ($10^{-1}$)&...&$5.4_{-0.7}^{+0.3}$\\
&norm ($10^{-3}$)&...&$1.4\pm0.3$\\
&F$_{\mathrm{unabs}}$&$10\pm3$&$9\pm3$\\
\hline
&$\chi^{2}$ (d.o.f.)&1232.9 (942)& 1180.4 (942)\\ 
\hline
\end{tabular}

\medskip
Note.---  Errors are reported at the 90\% confidence level and calculated from Markov chain Monte Carlo (MCMC) of chain length 500,000. The emissivity index $q$ and inclination $i$ were tied between {\sc relline} components, and the line emission is assumed to be isotropic. The outer disk radius was fixed at 990 $R_{g}$, the dimensionless spin parameter and redshift were set to zero for both the {\sc relline} and {\sc relxillNS} models. The temperature of the blackbody in the {\sc relxillNS} model was linked to the single temperature blackbody of the continuum. The {\sc relxillNS} model was set to reflection only and $f_{\mathrm{refl}}$ denotes the reflection fraction. The unabsorbed flux is taken in the $0.4-10.0$ keV band and given in units of $10^{-9}$ \fluxcgs. For reference, for $a=0$ and canonical NS mass $M_{\mathrm{NS}}=1.4\ M_{\odot}$, $1\ R_{\mathrm{ISCO}}=6\ R_{g}=12.4$ km.
\end{center}
\end{table}

As a further refinement, we replace the Gaussian line components with the relativistic reflection line model {\sc relline} \citep{dauser10}. 
We fixed the spin parameter, $a=cJ/GM^{2}$, to $a=0$ since most NS in LMXBs have $a\leq0.3$ (\citealt{galloway08}; \citealt{miller11}). The choice of spin does not have a strong impact on our results; the difference in position of the $R_{\mathrm{ISCO}}$ between $a=0.0$ and $a=0.3$ is less than 1 $R_{g}$ (where $R_{g}=GM/c^{2}$). We allowed the inner disk radii to be independent for each line component but tied the inclination and emissivity index between the lines. The outer disk radius was fixed to 990 $R_{g}$ in each case. The double relativistic reflection provides a $10\sigma$ improvement in the overall fit, although a Gaussian line cannot be statistically ruled out for the Fe~L complex. We find \rin\ of $1.4_{-0.1}^{+0.2}$ $R_{\mathrm{ISCO}}$ for Fe~L and $1.03_{-0.03}^{+0.13}$ $R_{\mathrm{ISCO}}$ for Fe~K. Additionally, the inclination of $10.1_{-0.2}^{+4.9}$ degrees from these lines agrees with the early \nustar\ results \citep{miller13} and optical spectroscopy \citep{cornelisse13}. Table 1 reports the values of all free parameters. Figure 3(a) shows the ratio of the data to the overall fit as well as the {\sc relline} model components. The increase in line energy for the Fe~L feature is due to the {\sc relline} model interpreting the feature as a single line rather than a complex of lines between 0.9 to 1.3 keV. There are still residuals in the $6-7$ keV energy band suggesting that {\sc relline} is unable to fully account for the Fe~K line, hence we proceed with applying a self-consistent reflection model.

\begin{figure} 
\begin{center}
\includegraphics[width=8.2cm]{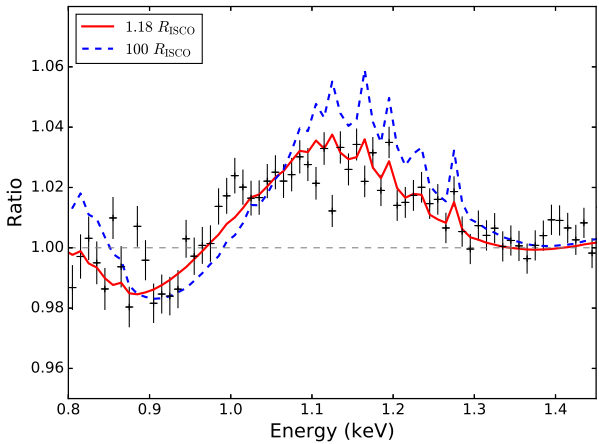}
\caption{Best-fit reflection model reported in Table 1 at 1.18 $R_{\mathrm{ISCO}}$ (red) and contrasting 100 $R_{\mathrm{ISCO}}$ (blue) overlaid on the \nicer\ data to highlight the broad Fe~L shell blend between 0.9 and 1.3 keV. The larger inner disk radius relaxes the relativistic effects to show the narrow emission lines in that region, likely due to Mg III-VII.
The data were rebinned for clarity.}
\label{fig:Felinemoratio}
\end{center}
\end{figure}

To obtain a more physical description of these features, we employ a preliminary version of the fully self-consistent reflection model, {\sc relxillNS}, which computes illumination of the disk by a blackbody spectrum (rather than the power-law input of the original {\sc relxill} model, \citealt{garcia14}). The model allows for the input blackbody temperature $kT_{bb}$,  log of the ionization parameter $\log \xi$, iron abundance $A_{Fe}$, and log of the density of the disk $\log (N$ [cm$^{-3}$]$)$. All other elements are hard-coded to solar abundance. In order for the model to pick up both the Fe~L blend and Fe~K line, the disk requires a high density and $\sim5$ times solar iron abundance (see \citealt{garcia16} and \citealt{ludlam17} for discussion of disk density and iron abundance).
This fit is reported in Table 1. Figure 3(b) shows the ratio of the data to the model as well as the reflection component. In Figure 4 we plot the reflection model in the region of the low-energy feature to demonstrate the blending of the Fe~L shell transitions. In order to illustrate the local-frame emission spectrum, which better shows the line complex features, we set $R_{in}$ to 100 $R_{\mathrm{ISCO}}$, a value so large as to effectively remove relativistic distortions.
The narrow emission lines in the same region as the broad Fe ~L shell are likely due to a lower-Z  element such as Mg~III--VII. 

The single inner disk radius inferred from the {\sc relxillNS} fit falls between the radii obtained from the Fe K line and Fe L blend in the {\sc relline} fit.
This could be due to the model applying the same physical conditions to each line when they could be arising from different locations and/or ionizations within the disk. We currently lack the data quality needed for a double {\sc relxillNS} fit to explore multiple ionization zones.
 
\section{Discussion}
Through the sensitivity and passband of \nicer\ we detected a broad Fe~L blend and Fe~K in the persistent emission of Ser X-1. We confirm that these lines are not native to \nicer\ instrumentation by comparing the spectra to observations made by \nustar\ and \xmm/RGS. \nicer\ captures $\sim1.34\times10^{6}$ photons in the Fe~L band and $\sim2.75\times10^{5}$ photons in the Fe~K band in just 4.5 ks. These lower-energy Fe~L lines have the potential to improve the statistical power of disk reflection and, ultimately, can be a very important tool for placing constraining upper limits on the radii of neutron stars if the lines indeed arise from the innermost regions of accretion disks. The position of the inner disk radius inferred from the Fe~L blend ($1.4_{-0.1}^{+0.2}$ $R_{\mathrm{ISCO}}$) is consistent with the value inferred from the Fe~K line ($1.03_{-0.03}^{+0.13}$ $R_{\mathrm{ISCO}}$) within the joint $3\sigma$ uncertainties. This is similar to the inner disk radius implied by the {\sc diskbb} normalization of the best fit model ($\sim21-28$ km) for an inclination of $10^{\circ}$, distance of $7.7\pm0.9$ kpc \citep{galloway08}, and color correction factor of 1.7 \citep{kubota98}. Our results for the position of the inner disk agree with previous spectral studies that utilize data from different observatories, such as {\it Suzaku}, {\it Chandra}, \xmm, and \nustar\ (\citealt{bs07}; \citealt{cackett08}, \citeyear{cackett10}; \citealt{chiang16}; \citealt{miller13}).

Furthermore, we demonstrate that both the low-energy blend and Fe~K line can be modeled with self-consistent reflection, but we are at the limits of the current data set.  As the mission progresses and calibration is improved, we will gain a larger sample of data from which we can explore a double {\sc relxillNS} model to determine the locations of these features individually and potentially explore the ionization structure of the disk \citep{ludlam16}. In a forthcoming paper, a final version of {\sc relxillNS} will be described and applied to a larger set of \nicer\ data, with the goal of further enhancing our understanding of the innermost disk around neutron stars.

\acknowledgements{R.M.L.\ acknowledges funding through a NASA Earth and Space Science Fellowship. A.C.F.\ acknowledges ERC Advanced Grant 340442. E.M.C.\ gratefully acknowledges NSF CAREER award AST-1351222.}\\

\end{document}